\documentclass[twocolumn,showpacs]{revtex4}
\usepackage{amssymb}
\usepackage{makeidx}
\usepackage{amsmath}
\usepackage{graphicx}
\usepackage{epstopdf}
\usepackage{dcolumn}
\usepackage{bm}

\begin{document}

\title{High harmonic generation in triangular graphene quantum dots}
\author{B.R. Avchyan}
\author{A.G. Ghazaryan}
\email{amarkos@ysu.am}
\author{K.A. Sargsyan}
\author{Kh.V. Sedrakian}
\affiliation{Centre of Strong Fields, Yerevan State University, 1 A. Manukian, Yerevan
0025, Armenia }

\begin{abstract}
Higher harmonic generation in plane graphene quantum dots initiated by intense coherent radiation is 
investigated, using dynamical Hartree-Fock mean-field theory. A microscopic theory describing the extreme 
nonlinear optical response of plane graphene quantum dots is developed. The closed set of differential equations 
for the single-particle density matrix at the graphene quantum dots-strong laser field 
multiphoton interaction is solved numerically. The obtained solutions indicate the significance 
of the type of edge and lateral size, and  bandgap and laser field strength in the high harmonic 
generation process on the triangular graphene quantum dot.

\end{abstract}

\date{\today }
\maketitle



\section{ Introduction}

In the last decade, there has been a growing interest to extend high harmonic generation (HHG) 
to two-dimensional (2D) crystals and nanostructures, such as semimetallic graphene \cite%
{aa}, and semiconductor transition metal dichalcogenidescite \cite{XX}. The role of graphene as 
an effective nonlinear optical material has been discussed in many theoretical \cite%
{1,2,3,4,5,6,7,8,9,10,11,12,13,14,15,girk,111,222}%
, and experimental \cite{16}, \cite{17} studies that consider various extreme nonlinear optical effects, in particular, 
HHG, which takes place in strong coherent radiation fields in the multiphoton regime 
at excitation of such nanostructures \cite{18}, \cite{19}. On the other hand, apart 
from the remarkable and unique electronic and optical properties of graphene, the lack 
of an energy gap as a semimetal greatly limits their applicability, 
in contrast, for example, to bilayer graphene \cite{20,21,22,23,24,25}.

The problem of a zero energy gap has been solved by decreasing the lateral size of graphene  \cite{bb}.
 As a result of dimensional quantization, an energy gap opens. Semimetallic graphene of finite dimensions becomes a semiconductor.
Among carbon nanostructures, there are of particular interest as a nonlinear medium the graphene ribbons (nanoribbons)
 \cite{bb}, \cite{cc}, graphene-like quantum dots \cite{1111, arxiv}, such as closed-convex fullerenes of 
different basic symmetry, and the graphene quantum dots (GQDs) of various lateral sizes.
The graphene nanostructure can be characterized by whether
 the sublattice symmetry is reserved. GQD has a gap that can be controlled by their lateral size,
 shape and type of edge \cite{bb}. GQDs behavior is quantitatively different for structures with zigzag and armchair edges,
 which is related to the edge states present in systems with zigzag edges \cite{ff}. So, it is of interest 
to investigate the HHG process in GQDs with different edges. Such nanostructure exhibits optical properties 
that are fundamentally different from those of graphene \cite {26,27,28}. 
At the same time, carriers in GQD have the same outstanding transport properties as in graphene \cite{1}.%

The important advantage of GQDs over the graphene nanoribbons \cite{ee} is the limitation 
of quasiparticles in in space. The latter can be crucial for the efficiency of HHG, since the space limitation prevents the propagation 
of the electron wave packet deposited into also one additional dimension and, 
therefore, can increase the HHG yield \cite{Lew}.
 
\begin{figure}[tbp]
\includegraphics[width=.35\textwidth]{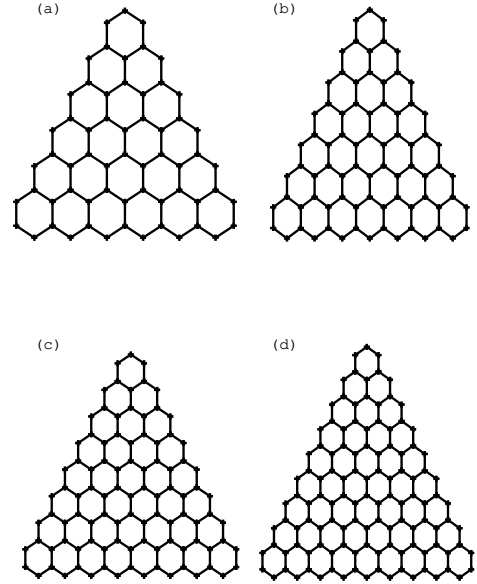}
\caption{ Schematic plot of the lattice of triangular QGD of zigzag edges with (a--d) $N=61, 78, 97, 118$ atoms, respectively. 
The distance between nearest neighboring atoms is $a\simeq1.42\ \mathring{A}$.}
\end{figure}

\begin{figure*}[tbp]
\includegraphics[width=.65\textwidth]{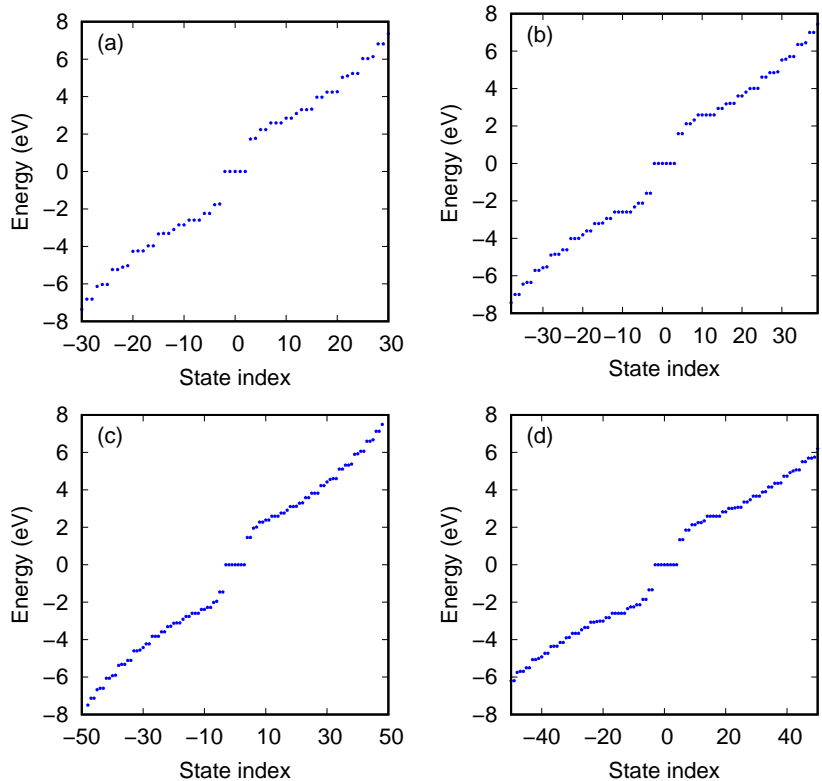}
\caption{The eigenenergies in triangular QGD of zigzag edges for (a--d) $N=61, 78, 97, 118$ atoms, respectively.}
\end{figure*}%
In the present work, the HHG in triangular GQDs caused by intense coherent radiation
 is investigated. The closed set of differential equations for the single-particle density 
matrix at the GQD-strong laser field multiphoton interaction is solved numerically in the scope of the microscopic theory,
 describing the extreme nonlinear optical response of GQDs is developed. GQD energy gap is 
controlled by its lateral size, shape, and type of edge. The obtained 
solutions indicate the significance of the lateral size on the HHG process 
in triangular GQDs type with either armchair or zigzag edge. Thus, we have investigated theoretically the effect of quantum confinement on HHG in GQDs 
by systematically varying the lateral size of a model dot.

The paper is organized as follows. In Sec. III, the set of equations for the single-particle density
matrix is formulated. In Sec. III, we consider multiphoton excitation and generation of harmonics in triangular GQD of the different type of edges and lateral size. 
Finally, conclusions are given in Sec. IV.%
 
\section{Dynamical Hartree-Fock mean-field theory for the high-harmonic
generation in QGD}

Let a plane GQD placed in the $xy$ plane bounded along the $x$, $y$ axes, that interacts with 
a plane quasimonochromatic electromagnetic (EM) wave. We will consider EM wave propagates perpendicular to the $xy$ plane. 
We assume neutral plane GQDs, which will be described 
using the empirical tight-- binding (TB) model firstly introduced in \cite{Wal}. Using of the TB model for 
GQDs were discussed in \cite{bb}, where the interball hopping is much smaller than
the on-ball hopping. TB Hamiltonian can describe finite-size
systems by restricting the tunneling matrix elements $t_{ij}$ to atoms within the quantum
dot. The total Hamiltonian can be written as: 
\begin{equation}
\widehat{H}=\widehat{H}_{\mathrm{0}}+\widehat{H}_{\mathrm{int}}.  \label{H1}
\end{equation}%
Here%
\begin{equation}
\widehat{H}_{0}=-\sum_{\left\langle i,j\right\rangle \sigma
}t_{ij}c_{i\sigma }^{\dagger }c_{j\sigma }+\frac{U}{2}\sum_{i\sigma }\left(
c_{i\sigma }^{\dagger }c_{i\sigma }-\frac{{n }_{i}}{2}\right) \left( c_{i\overline{%
\sigma }}^{\dagger }c_{i\overline{\sigma }}-\frac{{n }_{i}}{2}\right)  \label{Hfree}
\end{equation}%
is the free Hamiltonian of the GQD for TB model, where $c_{i\sigma }^{\dagger }$ creates an
electron with spin polarization $\sigma $ at site $i$, and $\left\langle
i,j\right\rangle $ indicates a summation over nearest neighbor sites
with the transfer energy $t_{ij}$, $ {n }_{i}$ -- average electron density for a site $i$.%
\begin{figure}[tbp]
\includegraphics[width=.35\textwidth]{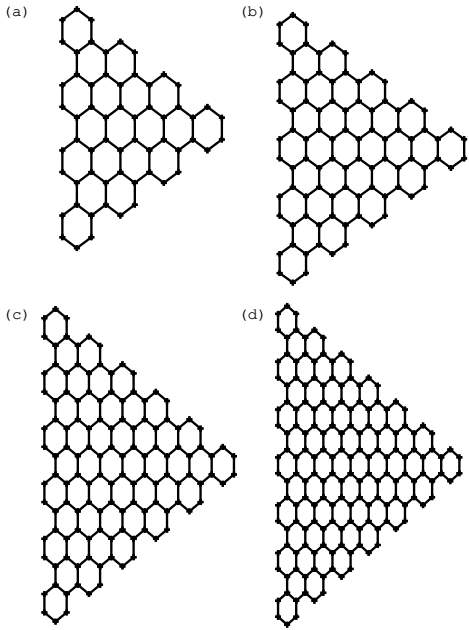}
\caption{The same as for Fig. 1 but for triangular QGD of armchair edges. (a--d) respectively correspond to $N=60, 90, 126, 168$ atoms. }
\end{figure} %
\begin{figure*}[tbp]
\includegraphics[width=.65\textwidth]{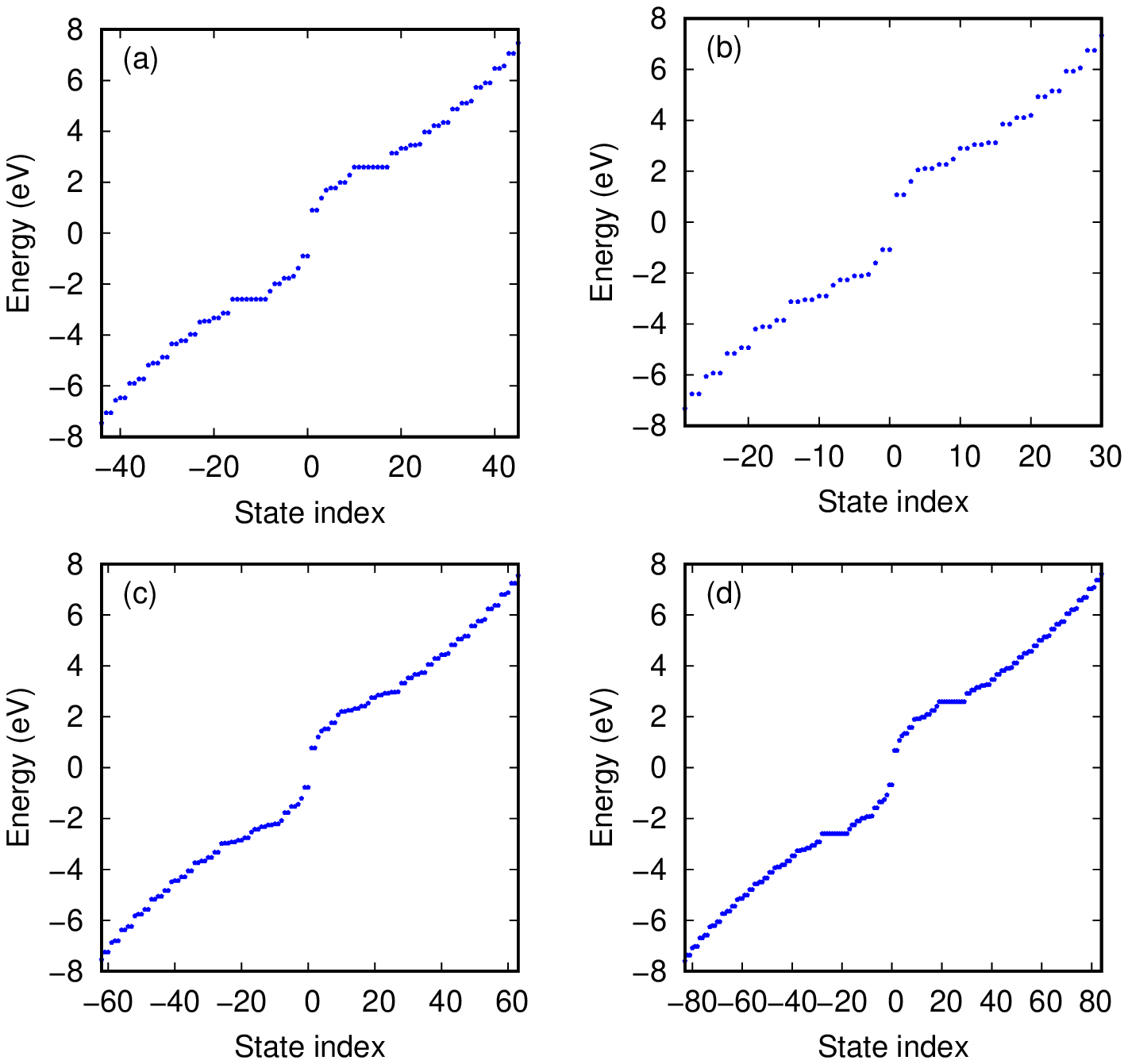}
\caption{The same as for Fig. 2 but for triangular QGD of armchair edges. (a--d) respectively correspond to $N=60, 90, 126, 168$ atoms. }
\end{figure*}%
The nonzero matrix elements of the TB Hamiltonian
given by the first term in (\ref{Hfree}) correspond to tunneling matrix element $t_{ij}$ between energy states on
neighboring sites.  The second term in (\ref{Hfree}) is the
electron-electron interaction (EEI) Hamiltonian ($\widehat{H}_{ee}$) in the
Hubbard approximation where all Coulomb scattering matrix elements are
neglected except for the on-site interaction terms $\sim U$ between spin-up
and down electrons occupying the same site $i$ ($\overline{\sigma }$ is the
opposite to $\sigma $ spin polarization). 
These operators $c_{i\sigma }^{\dagger }$, $c_{i\sigma }$  satisfy anticommutation rules: $\{ c_{i\sigma }c_{j\sigma }\} $= 
 $\{ c_{i\sigma }^{\dagger }c_{j\sigma }^{\dagger }\}=0$, and 
$\{ c_{i\sigma^{' } }c_{j\sigma }^{\dagger }\}$=$\delta_{ij}\delta_{ \sigma  \sigma^{' }}$, which guarantee antisymmetry of many-particle states.%

We assume, that before the interaction: $\left\langle c_{i\sigma }^{\dagger
}c_{i\sigma }\right\rangle ={n }_{i}$, $ {n }_{i}$ -- total electron density for a site $i$.  In the calculations, light-matter
interaction is described in the lenght-gauge via pure scalar potential:
\begin{equation*}
\widehat{H}_{\mathrm{int}}=e\sum_{i\sigma }\mathbf{r}_{i}\mathbf{E}\left(
t\right) c_{i\sigma }^{\dagger }c_{i\sigma },
\end{equation*}%
with the elementary charge $e$, position vector $\mathbf{r}_{i}$, and the
electric field strength $\mathbf{E}\left( t\right)$. In the Hamiltonian we
neglect the lattice vibrations. The hopping integral $t_{ij}$ between nearest-neighbor
atoms of QGDs can be determined experimentally,
and is usually taken to be $t_{ij}=2.7\ \mathrm{eV}$ \cite{1}.
The wave is assumed:%
\begin{equation}
{E}_{x}\left( t\right) =f\left( t\right) E_{0}\cos \omega
t\cos \theta, 
{E}_{y}\left( t\right) =f\left( t\right) E_{0}\cos \omega
t\sin \theta,  \label{field}
\end{equation}%
with the frequency $\omega $,
pulse envelope $f\left( t\right) =\sin ^{2}\left( \pi t/\mathcal{T}\right) $, the angle $\theta $  to 
the $x$ axis. The pulse duration $\mathcal{T}$ is taken to be $20$ wave cycles: $%
\mathcal{T}=40\pi /\omega $. From Heisenberg equation $i\hbar \partial 
\widehat{L}/\partial t=\left[ \widehat{L},\widehat{H}\right] $ one can
obtain evolutionary equations for the single particle density matrix $\rho
_{ij}^{\left( \sigma \right) }=\left\langle c_{j\sigma }^{\dagger
}c_{i\sigma }\right\rangle $. In addition we will assume that the system
relaxes at a rate $\gamma $ to the equilibrium $\rho _{0ij}^{\left( \sigma
\right) }$. distribution. To obtain closed set of equations for the single
particle density matrix $\rho _{ij}^{\left( \sigma \right) }=\left\langle
c_{j\sigma }^{\dagger }c_{i\sigma }\right\rangle $, EEI will be considered
under the Hartree-Fock approximation: 
\begin{equation}
\widehat{H}_{ee}^{HF}\simeq \frac{U}{2}\sum_{i\sigma }\left( \left\langle
c_{i\sigma }^{\dagger }c_{i\sigma }\right\rangle -\frac{{n }_{i}}{2}\right) \left(
c_{i\overline{\sigma }}^{\dagger }c_{i\overline{\sigma }}-\frac{{n }_{i}}{2}\right)
.  \label{HFU}
\end{equation}%
Thus, we obtain the following equation for the density matrix: 
\begin{equation*}
i\hbar \frac{\partial \rho _{ij}^{\left( \sigma \right) }}{\partial t}%
=\sum_{k}\left( t_{kj}\rho _{ik}^{\left( \sigma \right) }-t_{ik}\rho
_{kj}^{\left( \sigma \right) }\right) +U\left( \rho _{ii}^{\left( \overline{%
\sigma }\right) }-\rho _{jj}^{\left( \overline{\sigma }\right) }\right) \rho
_{ij}^{\left( \sigma \right) }
\end{equation*}%
\begin{equation}
+e\mathbf{E}\left( t\right) \left( \mathbf{r}_{i}-\mathbf{r}_{j}\right) \rho
_{ij}^{\left( \sigma \right) }-i\hbar \gamma \left( \rho _{ij}^{\left(
\sigma \right) }-\rho _{0ij}^{\left( \sigma \right) }\right) .  \label{evEqs}
\end{equation}%

\begin{figure*}[tbp]
\includegraphics[width=.65\textwidth]{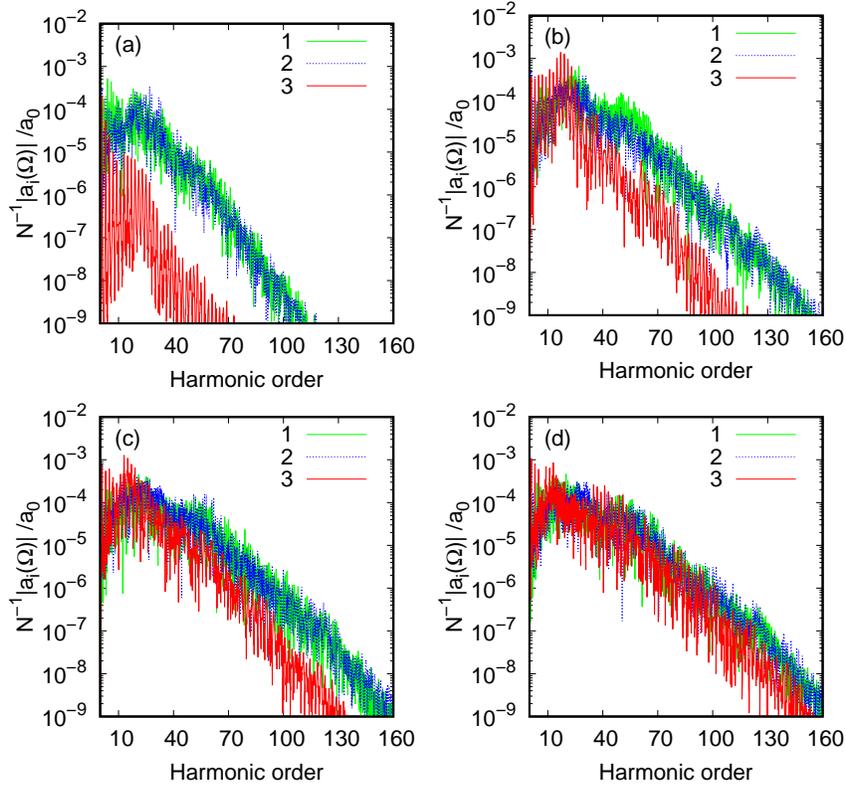}
\caption{(Color online) The HHG emission rate represents via dipole acceleration Fourier transformation  $%
N^{-1}|a_{i}\left( \Omega \right)|/a_{0}$ in the logarithmic scale versus the harmonic number for triangular QGD. (1) presents the components $%
|a_{x}|$ with zigzag and (3) armchair edges; (2)   $%
|a_{y}|$ with zigzag edges. (a--d) for (1, 2) correspond to $N=61, 97, 118, 141$ atoms, and (a--d) for (3) correspond to $N=60, 90, 126, 168$ atoms.
 The EM wave is assumed to be linearly polarized along $x $ axis. The wave frequency is $\protect\omega =0.1\ \mathrm{eV}/\hbar $ and 
field strength is $E_{0}=0.3\ \mathrm{V/\mathring{A}}$.
The spectra are shown for moderate (typical) EEI energy: $U=3\ \mathrm{eV}$. The relaxation
rate is $\hbar \protect\gamma =50\ \mathrm{meV}$.}
\end{figure*}%

\begin{figure*}[tbp]
\includegraphics[width=.75\textwidth]{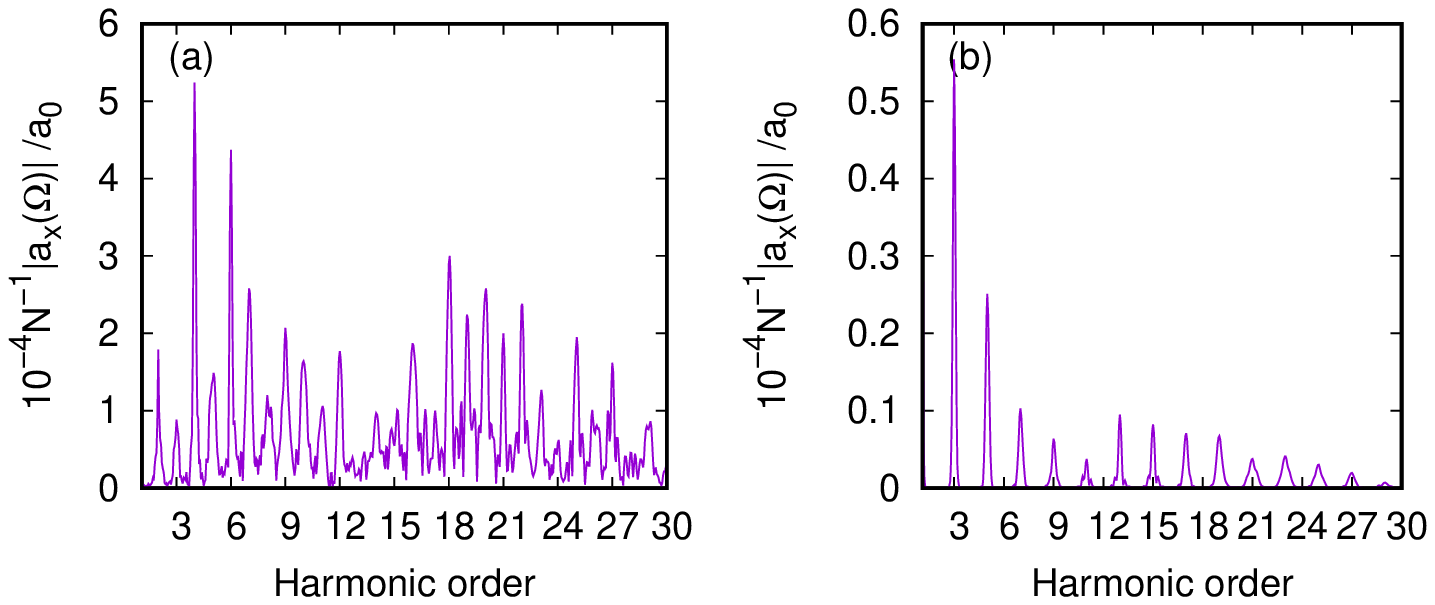}
\caption{(Color online) The same as for Fig. 5 but without the logarithmic scale. (a) shows harmonics (even and odd) in 
a triangular GQD with zigzag edges for $N=60$ atoms, (b) -- harmonics (only odd) in a GQD with armchair edges for $N=61$ atoms.}%
\end{figure*}%

We numerically diagonalize the tight-binding Hamiltonian $\widehat{H}_{0}$.
For a half-filled system, the static Hartree-Fock Hamiltonian vanishes $%
\widehat{H}_{ee}^{HF}\simeq 0$. It should be mentioned that EEI in
Hartree-Fock limit is included in empirical hopping integral between
nearest-neighbor atoms $t$ which is chosen to be close to experimental
values. Thus, on-site EEI in the Hartree-Fock approximation is relevant for
the quantum dynamics initiated by the pump laser field and as we will see
below considerably modifies the HHG spectrum. With the numerical
diagonalization, we find eigenstates $\psi _{\mu }\left( i\right) $ and
eigenenergies $\varepsilon _{\mu }$ ($\mu =0,1..N-1$). The results of
numerical diagonalization are shown in Fig. 2, 4. %
Without tunneling all energy levels
were degenerate,  $\psi _{\mu }\left( i\right) =0$ . So, the tunneling removed the degeneracy and led
to the formation of the band of 1--2 valence states below the Fermi level, a band of 1--2
conduction states above the Fermi level, $\varepsilon _{\mu }=0$ , and a gap across the Fermi level (see also  \cite{bb}). The quantum dynamics of
strong field periodically driven QGD is governed by a closed
set of differential equations (\ref{evEqs}), which should be solved with the
proper initial conditions. We construct initial a density matrix $\rho
_{0ij}^{\left( \sigma \right) }$ via the filling of electron states in the
valence band according to the Fermi--Dirac-distribution. Since the energy
gap is large enough we assume Fermi--Dirac-distribution at zero temperature:%
\begin{equation*}
\rho _{0ij}^{\left( \sigma \right) }=\sum_{\mu =N/2}^{N-1}\psi _{\mu }^{\ast
}\left( j\right) \psi _{\mu }\left( i\right) .
\end{equation*}%

\section{Numerical results for HHG efficiency in triangular QGDs}

The HHG spectrum is evaluated from the Fourier transformation $\mathbf{a}%
\left( \Omega \right) $ of the dipole acceleration $\mathbf{a}\left(
t\right) =d^{2}\mathbf{d}/dt^{2}$. The dipole is defined as $\mathbf{d}%
\left( t\right) =\left\langle \sum_{i\sigma }\mathbf{r}_{i}c_{i\sigma
}^{\dagger }c_{i\sigma }\right\rangle $. For convince we normalize the
dipole acceleration by the factor $a_{0}=\overline{\omega }^{2}\overline{d},$
where $\overline{\omega }=1\ \mathrm{eV}/\hbar $ and $\overline{d}=1\ 
\mathrm{\mathring{A}}$. The power radiated at the given frequency is
proportional to $\left\vert \mathbf{a}\left( \Omega \right) \right\vert ^{2}$%
. In order to clarify the main aspects of HHG in triangular GQDs, we assume
the excitation frequency is $\protect\omega =0.1\ \mathrm{eV}/\hbar $, that is much smaller than the typical gap $U\simeq3\ 
\mathrm{eV}$. The relaxation rate is taken to be $\hbar \protect\gamma =50\ \mathrm{meV}$. 
For the most calculations, the wave (\ref{field}) is assumed to be linearly
polarized along the $x $ axis ($ \theta =0$). The $ x $ axis is in the plane of Figs. 1, 3, and is directed along $ x $ to the right.%

\begin{figure*}[tbp]
\includegraphics[width=.65\textwidth]{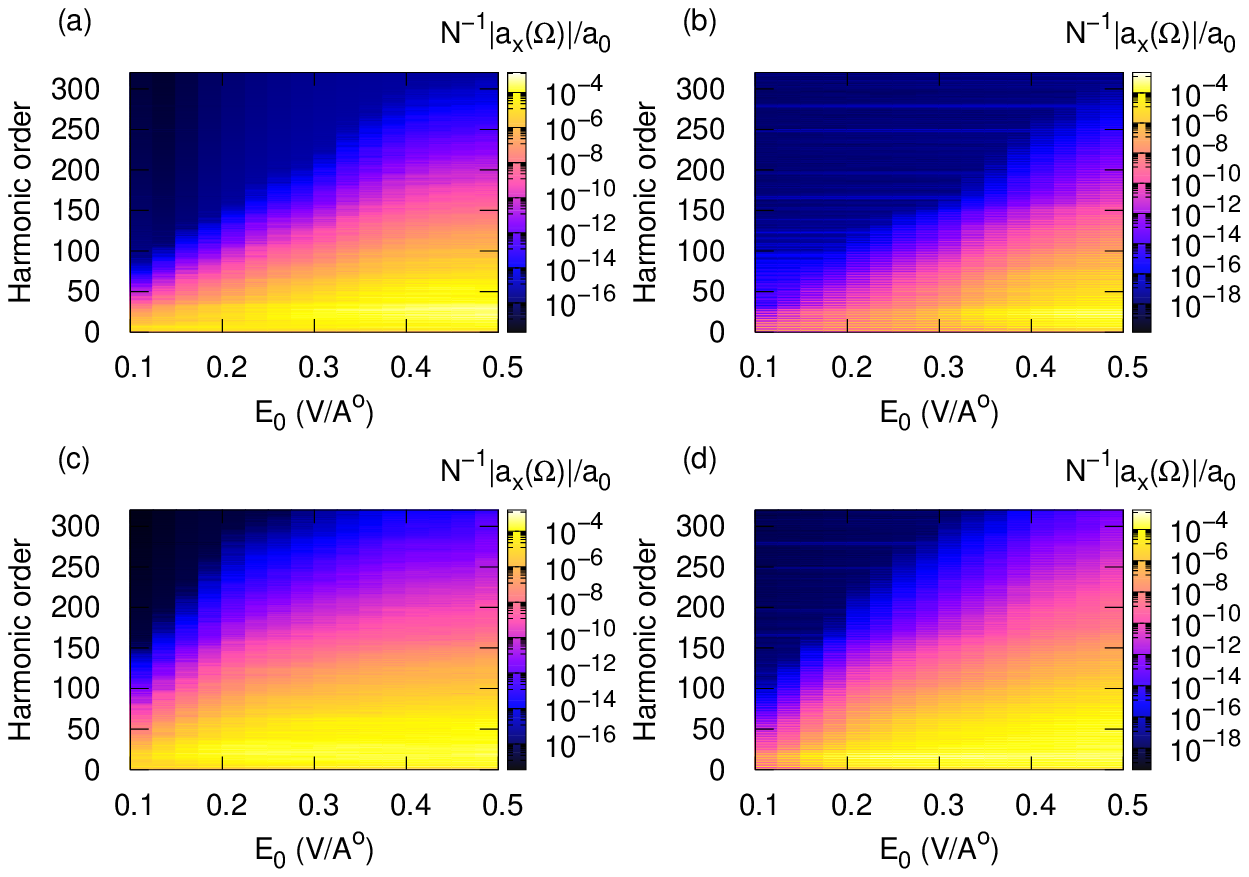}
\caption{(Color online) The color bar
represents the HHG emission rate via dipole acceleration Fourier transformation $%
N^{-1}|a_{x}\left( \Omega \right)|/a_{0}$ in strong field regime in the logarithmic scale versus hurmonic order and  EM
 field amplitude for triangular QGD of different edges and carbon atom numbers.  
(a, c) correspond to atom numbers $N=61$ and $N=118$ for QGD with zigzag edges, 
(b, d) correspond to  atom numbers $N=60$ and $N=126$ for QGD with armchair edges.
 The wave is assumed to be linearly polarized along  $x $ axis.
The wave frequency is $\protect\omega =0.1\ \mathrm{eV}/\hbar $ and the EEI energy: $U=3\ \mathrm{eV}$. 
The relaxation rate is $\hbar \protect\gamma =50\ \mathrm{meV}$. }
\end{figure*}%

\begin{figure*}[tbp]
\includegraphics[width=.65\textwidth]{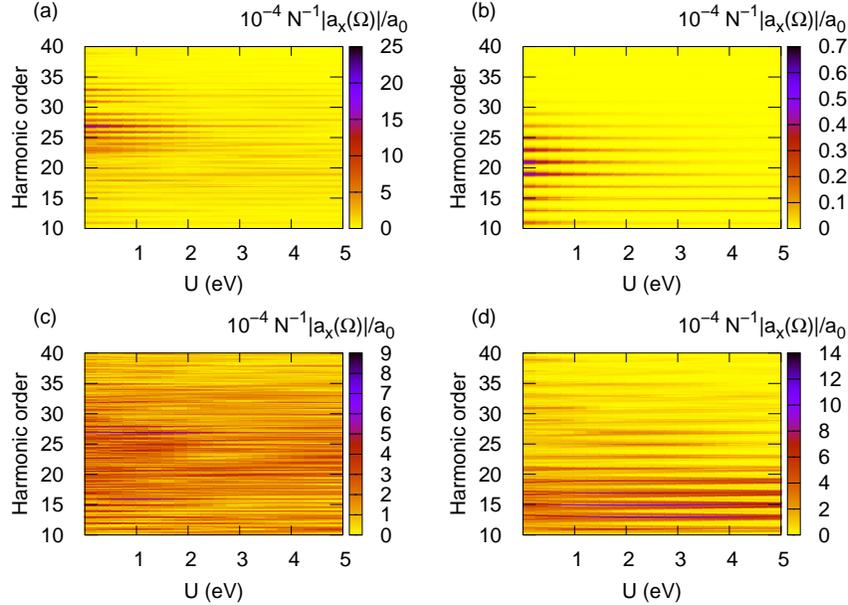}
\caption{(Color online) The same as in Fig. 7 but without the logarithmic scale, versus EEI energy and harmonic order at fixed EM field
strength $E_{0}=0.3\ \mathrm{V/\mathring{A}}$. }
\end{figure*}%

\begin{figure*}[tbp]
\includegraphics[width=.65\textwidth]{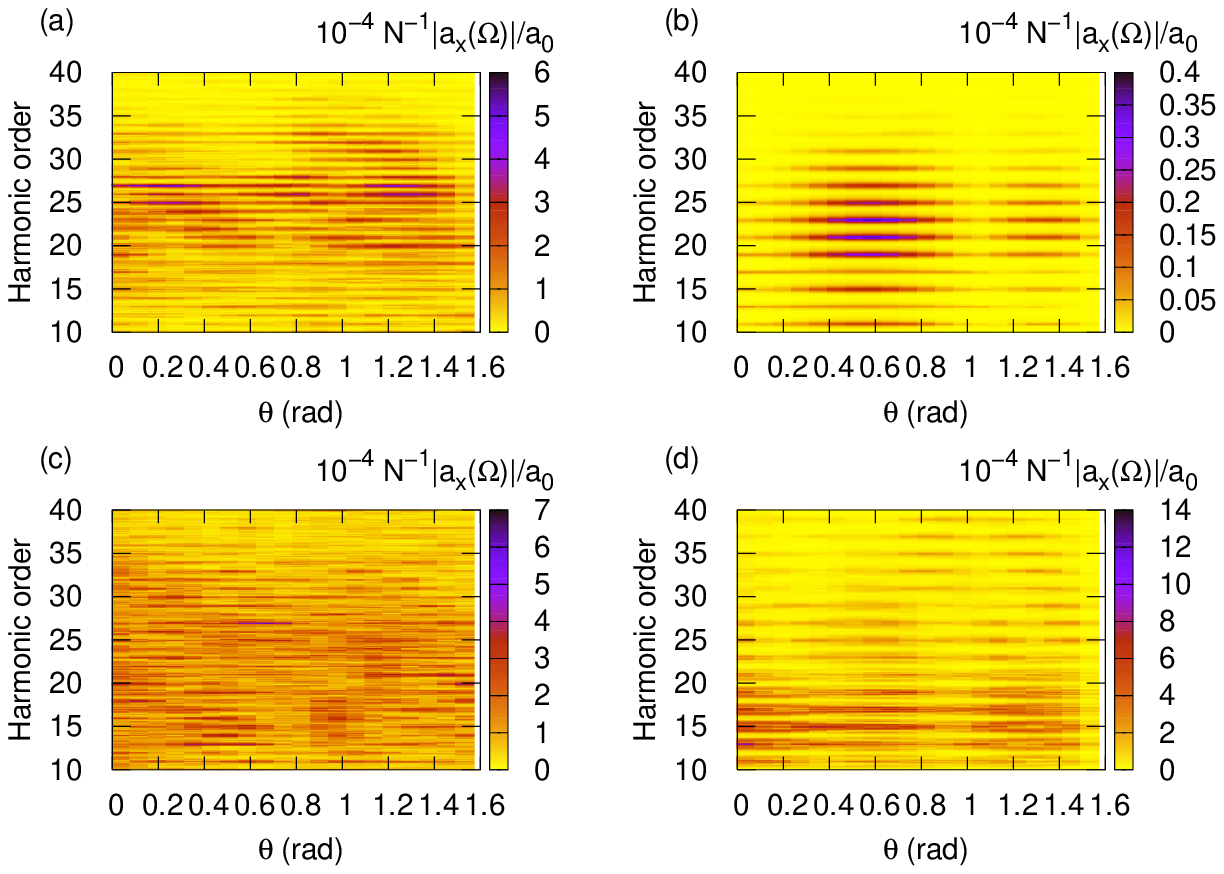}
\caption{(Color online) The HHG emission rate via dipole acceleration Fourier transformation$%
N^{-1}|a_{x}\left( \Omega \right)|/a_{0}$ versus hurmonic order and angle $ \theta$ of the intensity of the EM field with the $ x $ axis.
 (a, c) correspond to the number of atoms $ N = 61 $ and $ N = 118 $ for a triangular GQD with zigzag edges,
 (b, d) correspond to the number of atoms $ N = 60 $ and $ N = 126 $ for a triangular GQD with armchair edges. 
The wave frequency is $\protect\omega =0.1\ \mathrm{eV}/\hbar $, the field
strength is $E_{0}=0.3\ \mathrm{V/\mathring{A}}$, the EEI energy is $U=3\ \mathrm{eV}$. 
The relaxation rate is $\hbar \protect\gamma =50\ \mathrm{meV}$. }
\end{figure*}%

Figs. 1, 3 demonstrate schematically a graphene lattice, and Figs. 2, 4 show TB energy spectrum in the vicinity 
of the Fermi level, $\varepsilon _{\mu }=0$ , for a triangular GQD with zigzag and armchair edges for the different numbers of carbon atoms. 
As it is shown in Figs. 1--4, with an increase in the number of lattice atoms, the density of eigenstates increases. 
As will be seen later, this will increase the probabilities of multiphoton HHG. 

To compare the HHG radiation spectra in the triangular GQDs with different edges at different numbers of lattice atoms,  in further figures
we have plotted all results for the spectra via the normalized atom number $N$. In Fig. 5 plotted the components $ | a_ {x} \left (\Omega \right) | $
 and $ | a_ {y} \left (\Omega \right) | $ in the strong EM wave with the amplitude $ E_ {0} = 0.3 $ B $ \mathrm {/ \mathring {A}} $, 
and the EEI energy $U\simeq3\ \mathrm{eV}$. As shown in this figure, in a strong laser field, multiphoton harmonics are significant, and the HHG
 yields for GQDs are equally significant for both armchair and zigzag edges, especially for large $ N $. 
In both cases, the HHG spectra have a multistep plateau structure, which is associated with the excitations of energy
 eigenstates between the unoccupied energy levels and the occupied level \cite {bb}. GQD with armchair-like edges 
has axial symmetry (see Fig. 3), and, in particular, in this case $ | a_ {y} | = 0 $ (therefore, it is absent in Fig. 5), 
while in the case of zigzag edges both components $ | a_ {x} | $ and $ | a_ {y} | $ are essential (fig. 5). 
In addition, only odd harmonics are visible in the HHG spectrum in armchair-shaped GQDs, as in ordinary graphene \cite{aa}. 
However, for zigzag edges, due to the absence of inversion symmetry, both odd and even harmonics are present in the HHG emission spectrum. 
To show this clearly, in Fig. 6 separately shows the results for HHG for the first thirty harmonics with approximately the same number($N\simeq60$)
 of atoms of a triangular GQD with different edges.

 Next, we consider the HHG spectra as a function of pump wave intensity. Fig. 7 demonstrates
 the HHG spectra as a function of EM field amplitude and the
harmonic order for a fixed EEI energy $U=3$ $\mathrm{eV}$. To compare in Fig. 7a, c and Fig. 7b, d we will investigate 
the QGDs with a similar number of carbon atoms but different border edges. As shown in Fig. 7, 
the HHG probability increases either with an increase in the number  $N$ of dots' particles or with
 an arise more new energy states (also see Figs. 1--4). The cutoff harmonic linearly increases
with increasing the field strength.  Then, reaching the harmonic $n_{\mathrm{%
cut}}$ which corresponds to the transition of the lowest occupied energy state
to the highest unoccupied one,l the HHG rate is saturated (stepped yellow envelope). The harmonic cutoff number can be seen in Fig. 5: $n_{\mathrm{%
cut}}\simeq160$. Note that linear dependence of the cutoff harmonics on the field strength is
inherent to HHG via discrete levels, or in crystals with linear energy
dispersion. As in atomic cases for QGD with the increase of the
pump wave strength at a fixed photon energy the cutoff harmonic energy ($\hbar \omega N_{\mathrm{%
cut}}$) is increased.%

Besides, as was shown in \cite{22} the on-site EEI suppresses the charge fluctuation and reduces the absorbed
energy. Suppression of HHG yields due to EEI is also expected. The latter is
shown in Fig. 8, where the HHG spectra in the strong EM field regime versus photon number and 
EEI energy are shown for different border edges and carbon atom numbers. As can be seen from Fig. 8, for a small number of $ N$ GQD atoms
with an increase the EEE energy, the HHG rate is generally suppressed 
(therefore, Fig. 8 shows the dependence for mean harmonics with numbers $ <40 $). The latter is not the case
for large $ N$, when the density of energy states increases (Fig. 8c, d). The HHG spectrum ceases to depend on the Coulomb EEE.
 This property is inherent in ordinary graphene \cite {aa}, unlimited in space.%

We also investigated the HHG spectra dependence versus the EM wave strength orientation.
In Fig. 9 we plot the HHG spectra versus the photon number and pump wave strength orientation under $x $ axis for a given strength 
($E_{0}=0.3\ \mathrm{V/\mathring{A}}$) and frequency and for moderate EEI energy $U=3$ $\mathrm{eV}$. As can be seen from 
Fig. 9, the orientation of the pump wave at different angles to the $ x $ axis leads to different harmonic spectra. 
This is due to the fact that the triangular GQD has no inversion symmetry (see Fig. 1-4). As shown in fig. 9a, b for angles $ 0 <\theta <\pi / 2 $ 
the rate of middle harmonics (maxima for numbers $ \simeq10-40 $) increases, and higher harmonics are suppressed.
 However, in Fig. 9c, d, it can be seen that with an increase in the density of energy states, this regularity
is violated, and the HHG spectrum ceases to depend on the orientation of the EM wave force. By this, the GQD becomes similar to graphene unbounded in space \cite {aa}.%

\section{Conclusion}

We have studied the influence of intense coherent radiation on GQDs. The microscopic theory has been developed to describe the 
extreme nonlinear optical response of the triangular GQDs. A closed system of differential equations 
for a one-particle density matrix in the multiphoton interaction of a GQD 
with a strong laser field is solved numerically. The solutions obtained indicate the importance 
of the type of edge and lateral size, as well as the significance of the bandgap and the magnitude of the laser field for the HHG process in plane GQDs. 
The harmonic cutoff number increases linearly with increasing field strength. As in the case of HHG on atoms, for a GQD, with an increase 
the pump wave strength at fixed photon energy, the harmonic cutoff energy ($\hbar \omega n_{\mathrm{%
cut}}$) increases linearly. Due to the absence or presence of symmetry of the sublattice in a triangular GQD
 with zigzag edges, harmonics of both odd and even order appear during the generation in the field of an EM wave, 
when only odd harmonics are significant for armchair edges. We also investigated the HHG spectra dependence versus the pump wave strength orientation.  
As the numerical results show, due to the difference in the symmetry of the sublattice, the same angles give different partial yields in the 
HHG spectra for triangular GQDs with armchair and zigzag edges. In addition, the rate of middle harmonics for a small number of atoms increases, 
while higher harmonics are suppressed, which is not the case with an increase in the density of energy
states. In addition, with an increase in the density of energy states, the rule is violated when, in the case of a small number of GQD atoms, 
with an increase in the EEI energy, the HHG rate is generally suppressed. Thus, with an increase in the average
the density of eigenstates of the GQD behaves like graphene unbounded in space. So, we investigated the size and shape of the GQD using quasiparticle confinement
 in space. The results obtained show that GQDs can serve as an effective medium for the generation of even and odd high-order 
harmonics when interacting with a laser field of moderate intensity due to the limitation of quasiparticles in the GQD. 
In addition, the HHG probability increases with an increase in the number of GQD atoms or with the appearance of new energy states. 
This is a potential way to increase the quantum yield and photon energy during HHG in graphene-like quantum dots.

\begin{acknowledgments}
The authors are deeply grateful to prof. H. K. Avetissian and Dr. G. F. Mkrtchian for permanent discussions and valuable recommendations.
The work was supported by the  Science Committee of RA in frames of the
research project  20TTWS--1C010.
\end{acknowledgments}

\end{document}